# Ultra-conformable Liquid Metal Particle Monolayer on Air/water Interface for Substrate-free E-tattoo

*Fali Li, Wenjuan Lei, Yuwei Wang, Xingjian Lu, Shengbin Li, Feng Xu, Zidong He, Jinyun Liu, Huali Yang, Yuanzhao Wu, Jie Shang, Yiwei Liu*, Run-Wei Li**

F. Li, W. Lei, Y. Wang, X. Lu, S. Li, F. Xu, Z. He, J. Liu, H. Yang, Y. Wu, J. Shang, Y. Liu, R-W. Li
CAS Key Laboratory of Magnetic Materials and Devices, Ningbo Institute of Materials Technology and Engineering, Chinese Academy of Sciences, Ningbo, 315201, P. R. China.
E-mail: liuyw@nimte.ac.cn (Dr. Y. Liu); runweili@nimte.ac.cn (Prof. Dr. R.-W. Li)
F. Li, W. Lei, Y. Wang, X. Lu, S. Li, F. Xu, Z. He, J. Liu, H. Yang, Y. Wu, J. Shang, Y. Liu, R-W. Li
Zhejiang Province Key Laboratory of Magnetic Materials and Application Technology, Ningbo Institute of Materials Technology and Engineering, Chinese Academy of Sciences, Ningbo, 315201, P. R. China.
X. Lu, F. Xu, Z. He, J. Liu,Y. Liu, R-W. Li
College of Materials Science and Opto-Electronic Technology, University of Chinese Academy of Sciences, Beijing, 100049, P.R. China.

**Abstract:**

Gallium-based liquid metal is getting increased attention in conformal flexible electronics for its high electrical conductivity, intrinsic deformability and biocompatibility. A series of flexible devices are developed based on the micro-particles of liquid metal. But it is still challenging to fabricate conformal liquid metal film with a large area and high uniformity. Interfacial self-assembly is a competitive candidate method. Traditional interfacial self-assembly methods have difficulties assembling liquid metal particles because the floating state of the high-density microparticles could be easily disturbed by gravity. Here, we realized the multi-size universal self-assembly (MUS) for liquid metal particles with various diameters (0~500μm). By introducing a simple z-axis undisturbed interfacial material releasing strategy, the interference of gravitational energy on the stability of floating particles is avoided.

Benefits from this, the ultra-conformable monolayer film, with large area (>100 $cm^2$) and high floating yield (50%~90%), can be fabricated by liquid metal particles. Furthermore, the monolayer can be conformally transferred to any interesting complex surface such as human skin and plant leaf, to fabricate substrate-free flexible devices. Without interference from the mechanical response of traditional substrate, the liquid metal e-tattoo is more user-friendly and can realize feel-less continuous monitoring.

## 1. Introduction

With the increasing demand for integrating electronic devices into the human body, the conformity is becoming more crucial to ensure reliable operations.[1-3] As the surface of the human body is non-developable and experiences complex deformation in daily life, flat flexible devices have to work in a complex mechanical environment on human skin, resulting in instability and discomfort.[4-6] Tattoo-like electronic devices are becoming the trend has excellent comfort and user-friendliness[7,8]. Benefit from the intrinsic deformability[9,10] and outstanding electrical properties,[11,12] gallium-based liquid metals are promising candidates for fabricating conformal electronics. Based on liquid metal particles,[13,14] a variety of flexible devices have been developed for detecting physiological signals of the human body, and the conformity of devices is also improving.[15,16] Although widely used in flexible electronics, it is still challenging to fabricate large areas of a uniform conformal liquid metal film.

Interfacial self-assembly is a competitive candidate for developing conformal liquid metal film. It is a popular method in many fields for fabricating various nanomaterials-based monolayers. It can be applied to materials from metallic nanoparticles and nanowires,[17,18] carbon-based nanomaterials[19] to organic nanomaterials[20]. Compared with conventional conformal devices fabricating strategies, such as substrate sacrificial transferring, the interfacial self-assembled film is more amiable to complex surfaces.[21-24] Consequently, applying this method to preparing conformal liquid metal film is attractive.

However, traditional interfacial self-assembly methods are mainly applicable to nano-sized materials,[25-28] it has difficulties in assembling liquid metal particles, which are relatively large-size (100nm~500μm) and high density (~6g/$cm^3$). Theoretically, the surface tension of water is large enough to support liquid metal microparticles at the air-water interface [29]. Still, the balance can be easily broken by the kinetic energy accumulated during the interfacial self-assembly process.

Here we propose a multi-size universal self-assembly (**MUS**) strategy for fabricating liquid metal particles monolayer (**LMPM**). The monolayer can be conformally transferred from the air-water interface to arbitrary substrates. To reduce the interference factors in traditional methods, a z-axis undisturbed material-releasing strategy is introduced. Under the action of the Marangoni force, liquid metal particles can diffuse to the interface when the slurry of liquid metal particles and alcohol contacts the air/water interface. In this process, there are no interference factors in the vertical direction so that the liquid metal particles can float and assemble more efficiently in the air/water interface. Benefiting from this improvement, the MUS demonstrates its universality in various sizes, from several nanometers to hundreds of microns. This will pave the road toward a general and more effective method for obtaining large-area and uniform conformal liquid metal film, and further fabricating it into large-area conformal devices.

## 2. Results and discussion

Theoretically, the stability of particles on the interface is maintained by the kinetic energy barriers.[29] according to Young's equation (**Equation 1**). The contact angle for a particle at equilibrium depends on the interfacial tensions at particle-air, $\gamma_{pa}$, particle-liquid, $\gamma_{pl}$, and air-liquid, $\gamma_{al}$. If the particle detached from the interface and is immersed into the liquid phase, the free energy $\Delta G$ to be overcome to submerged is shown in **Equation 2**.[29] In the traditional assembling process, the droplet of colloidal falls at a specific height h. Here we can define a parameter $X_{energy}$ in **Equation 3**, which is the ratio of free energy ($\Delta G$) and kinetic energy ($\Delta E$) accumulated in the falling process. Taking the liquid metal micro-particles floating on water as an example, we can get **Equation 4**.

$$\cos\theta = \frac{\gamma_{pa}-\gamma_{pl}}{\gamma_{al}} \quad (1)$$

$$\Delta G = \pi r^2 \gamma_{al}(1-\cos\theta)^2 \quad (2)$$

$$X_{energy} = \frac{\Delta G}{\Delta E} = \frac{3\gamma_{al}(1-\cos\theta)^2}{4r\rho g h} \quad (3)$$

$$\text{For liquid metal:} \quad X_{energy} \leq \frac{1.75\times10^{-5}m}{r} \quad (4)$$

Consequently, a larger particle yields a smaller value of $X_{energy}$, thus decreasing particle stability in the assembling process. For metallic particles, it's easier to be disturbed by the kinetic energy when the radius exceeds a specific value (17.5 μm for liquid metal), thus making it difficult to assemble into a film at the air-liquid interface. $X_{energy}$ is the crucial parameter that determines the practicability of the interfacial assembly for large-size materials. In the MUS process, we eliminated the pre-step of uniformly dispersing the particles, and instead directly started from the micro-particles' precipitate in ethanol. The micro-particles in precipitation are released horizontally when contacting the air-liquid interface, and the Marangoni force mediates the process. Therefore, by decreasing the particles' kinetic energy in the vertical direction, this method can reduce the $X_{energy}$ and thus enable the interfacial self-assembly of liquid metal micro-particles.

To demonstrate the advantages of MUS strategy in detail, a comparative analysis is presented in **Figure 1**b-d. Firstly, for the nanomaterials in the traditional interfacial self-assembly method (Fig. 1b), $X_{energy} \ll 1$, and it can be easily dispersed uniformly in the ethanol to form a colloid. Adding the colloid dropwise to the water can quickly obtain a monolayer at the air-liquid interface. In contrast, as shown in Fig. 1b, micro-scale metallic materials are difficult to disperse stably in ethanol. In the process of adding this suspension to the water, the relation $X_{energy} \ll 1$ is no longer holds. Therefore, the microparticles are unstable at the gas-liquid interface. Disturbance such as kinetic energy accumulated during the falling process of droplets, can easily break this balance and cause the material to sink into the water. Taking the above considerations into account, the MUS is shown in Fig. 1d. This method does not need to disperse the material uniformly in ethanol, and can directly utilize the precipitation of particles. Placing the residue at the air-water interface can induce Marangoni forces due to the difference in surface tension between ethanol and water [14]. The Marangoni force can pull the particles out of the packed precipitate, and spread out at the interface. Since the force is parallel to the in-plane direction, and the interference factor (gravity) in the vertical direction is effectively evaded, the micro-particles are relatively stable and do not sink. The micro-particles are closely packed at the interface to form a monolayer.

The schematic diagram of the MUS is shown in Fig. 1a, which includes the precipitate preparation, interfacial self-assembly, conformal transfer, and film processing. Firstly, the liquid metal particles are stirred sufficiently in the ethanol, and liquid metal particles slurry is obtained after standing for minutes. Further, the slurry is transferred to make contact with the air-water interface. The Marangoni effect is induced by the difference of surface tension between the ethanol and water, which will cause Marangoni flow from the slurry to the interface. Consequently, the liquid metal particles diffuse along with the ethanol to the air-water interface and self-assembled into a monolayer. Furthermore, the particle film can be easily transferred to complex surfaces such as the human body. At last, a conformal liquid metal particle film is obtained, and the film can be used to prepare flexible circuits and sensors by methods such as mechanical sintering.

The MUS method exhibits a time-saving advantage in fabricating LMPM. **Figure 2**a-d shows the film preparation process. The film was prepared within the 20 s (diameter≈10 cm). In Fig.2a, a glass slide is used to carry the liquid metal slurry to contact the air-water interface. It can be seen that the particles form a flower-like pattern under the action of the Marangoni force. As the liquid metal particles diffuse into the air-water interface, the film shows up in the right of the beaker and saturates at 20 s (Fig. 2b-d). After removing the slurry, the LMPM covers the entire air-water interface.

The influencing factors on the film preparation are systematically studied in Fig.2e-j. The self-assembly process is greatly affected by particle size. However, the particle size of liquid metal obtained by standard methods (such as ultrasound, stirring, etc.) is relatively dispersed, which makes it challenging to study the relationship between particle size and the assembly process. Therefore, monodisperse liquid metal particles should be prepared firstly. The monodispersed liquid metal microparticles are obtained by the microtube spraying method (shown in Fig. S1a). High pressure is used to force liquid metal through a glass microtube, during which the liquid metal bulk shrinks to form microparticles. The diameter of the liquid metal particles can be easily controlled by controlling the diameter of the glass microtubes (shown in Fig. S1b).

In the previous theoretical analysis section, by defining the stability parameter $X_{energy}$, we predicted the vast impact of particle kinetic energy (vertical) on particle floating stability. Here, monodispersed liquid metal micro-particles (diameter≈30 μm) are taken as an example to experimentally demonstrate the importance of controlling the height of particle release. As shown in Fig. 2e, it is difficult to achieve particle flotation and assembly when releasing above or below the water surface. On the contrary, liquid metal particles can float and assemble efficiently by releasing liquid metal at the interface. For liquid metal particles with a size distribution of 10-700 μm, the flotation yields of the traditional method and the MUS method are compared in Fig. 2f. As the liquid metal microparticles is incompatible with the traditional method, almost no films are obtained at the interface. In contrast, the MUS method is effective for micron-sized liquid metal particles. For particles with diameters below 300 μm, the film yield is above 75%. Although this method improved a lot compared to the traditional way, it is still challenging to achieve 100%. Fig. S2 may provide some explanation for this. Here we can define two kinds of liquid metal particles, A and B, which are shown in Fig. S2, and both can diffuse into the interface from the slurry. B maintains a relatively constant speed and finally moves to the far side of the interface to assemble into a film, while A drops rapidly after a short distance. When the particle diameter exceeds 500 μm, the effect of gravity becomes prominent gradually, and the floating yield decreases with the diameter increase. For a size of 700 μm, even if the liquid metal particles can float on the interface by the MUS process. However, it is unstable and prone to falling into the water under gravity. Further, the method is not limited to the use of alcohol as the pre-dispersed phase of the particles. The effect of the different solutions on the MUS process was also tested (Fig. 2g). Unlike water-insoluble solvents (hexane), when using other water-soluble organic solvents (such as acetone, NMP, and DMF), it can also achieve high film yields (>75%).

As the melting point of liquid metal is close to room temperature, the MUS can also be affected by the temperature of the ambient environment. The time-lapse microscopic paragraph of the diffusion process at 0 °C is shown in Fig. 2h. At this temperature, the liquid metal particles are solidified. From the trajectory of a single particle, we can see no adhesion between particles during the material-releasing process. In contrast, when the temperature reaches 25 °C, the interior of particles presents a liquid state, and the oxide skin maintains its shape. Influenced by the adhesion of the oxide skin, the particles aggregate and are closely packed. The temperature also significantly influenced the floating yield of self-assembled particle film, which is shown in Fig. 2j. With increasing the temperature, the particles floating yields reduced. Therefore, for lower melt point materials, temperature plays an important role in the film yields when using the MUS method.

The morphology of LMPM based on liquid metal particles with different diameters is shown in **Figure 3**a-f. From the top view in Fig. 3a-c, it can be seen that the liquid metal particles are arranged in a close-packed state, and their diameters are about 150 μm, 40 μm, and 30 μm, respectively. The cross-sectional view in Fig. d-f shows that the film consists of a single layer of liquid metal particles whose thickness corresponds to the diameter of the individual particle. Except for liquid metal, we further explored the universality of the MUS method on other materials. Here we choose three representative materials, which are common metal (copper), magnetic material (FeSiB), and compound (ZnS). Fig.S3a–c shows the optical photos of the self-assembled particle films obtained by the MUS method for three kinds of particles, respectively. Fig.S3d-i shows the microscopic images of the above three microparticle films. From the SEM cross-sections in Fig.S3 g-i, it can be concluded that these films are composed of a single layer of particles, and the film thicknesses are all equal to the diameter of the individual particles.

The electrical properties of the liquid metal film are shown in Fig. 3g-h. The LMPM is obtained at the air-water interface based on mono-disperse liquid metal particles with an average size of 30 μm. Then, the film is transferred to the PDMS substrate, and the sample size is 1×4cm. The initial resistance of the sample fluctuates around $10^6$ Ω, as shown by the black line in Fig. 3g. The maximum strain for the first cycle is set at 20%. In the process of strain loading, the resistance is always in the $10^6$ Ω until the strain exceeds a certain threshold (about 18%), suddenly drops to near $10^5$ Ω, and remains relatively stable in the unloading process. In the subsequent tensile cycle, the maximum strain increases to 30%, 40%, and 50%, corresponding to the dark blue, blue, and light blue curves in Fig.3g, respectively. When the strain of the monolayer exceeds the highest strain in history, its resistance will begin to decline until the resistance reaches the order of 10 Ω. We speculate that this phenomenon is caused by particle breakage and conductive path formation during stretching.[30] The subsequent cyclic tensile test can also support this view (Fig. 3h). The sample is stretched for 1000 cycles with maximum strain at 50%. In the first cycle, the resistance rapidly decreases from the $10^6$ Ω to the magnitude of 10 Ω, and becomes relatively stable in the subsequent process. Inset shows the details of resistance changes after hundreds of stretching exercises. The initial resistance is 3.18 Ω, which decreases with the increase of strain and recovers with the unloading of strain.

The LMPM can be conformally transferred to complex surfaces, which is the base of conformal flexible electronic devices. Here we take the LMPM based on liquid metal particles of 30 μm as an example to demonstrate its conformal ability. Lift-off approach is adopted in the film transfer process. As shown in **Figure 4**a, the object (a volumetric flask) is first placed underwater and then pulled out of the water. During the lifting step, the LMPM automatically conformal covers the surface. The LMPM can be effectively transferred to flat objects (silicon wafers, Fig. 4b), and elastic spheres (silicone hemispheres, Fig. 4c). The deformability of the film on the PDMS substrate is shown in Fig. 4d. The film does not appear to fall off under 50% tensile strain. The film is further transferred to an uneven surface, such as the cherry leaf (shown in Fig.4e). Although the microstructure of the leaf is complex, the liquid metal particle film exhibits a conformal solid ability. Fig. 4f, g shows the SEM image of LMPM on a cherry leaf. It can be seen that the liquid metal particles can cover the undulating structures such as the veins of the leaf. In addition to uneven surfaces, liquid metal particle films can also be transferred to complex three-dimensional curved surfaces. The conformal attachment of LMPM to the surface of nitrile gloves is shown in Fig. 4h-i. From the front and back images of the glove, we can conclude that the entire surface has been completely covered with the liquid metal particle film. Compared with nitrile gloves, the surface of the human finger has a more complex wrinkle structure. Fig. 4j shows the LMPM transferred to the back of the finger, which is conformally attached to the skin. Also, as shown in the inset, the film can be transferred to the fingerprint, while the structure of fingerprint remains intact.

The conformal liquid metal particle film is essential for further obtaining conformal liquid metal-based flexible electronic devices. Fig. S5 demonstrates the patterning of LMPM and the circuit fabricating based on it. Firstly, the mask (made of paper) is attached to the object. Submerge the target object into the water, and the film will adhere to the surface of the object when lifted out of the water. After that, a brush is used to apply pressure on the film to active liquid metal particles and form a conductive path. At last, the patterning of the LMPM on the surface of the three-dimensional object can be realized after removing the mask.

The conformal LMPM depicts excellent potential as a conformal heater in skin electronics. Local heating of the lesion area is a common health care measure in daily life, and the key is achieving conformal contact with the lesion area. In **Figure 5**a, we used paraffin to simulate the raised structure of human skin lesions on the surface of the prosthesis, and then the conformal heating function of the LMPM is demonstrated on it. In fig. 5b, LMPM is conformally transferred to the surface of the lesion area. The projection of this area is a circle with a diameter of 1.5 cm, but its three-dimensional structure is relatively complicated. The inset of fig. 5b (side view) shows that the LMPM achieves conformal coverage of this region. It is worth mentioning that the film is straightforward to wash off after use. Fig. 5c shows the LMPM deposited on the joints of the author's hand. The film doesn't fall off during the normal movement of the human palm. But it can be easily washed off by soap in 1 minute (fig. 5d).

The conformal heating process is demonstrated in fig. 5e-g. A commercial low-voltage high-frequency inductive heating device is placed close to the LMPM. The high-frequency electric field will induce eddy currents in the metal film on the skin, thereby generating Joule heat. The temperature can be controlled by controlling the heating time and the distance between the induction coil and the skin. The infrared picture of the hand is shown in fig. 5f (heating for 10 seconds). The temperature at the position of the blue curve in fig. 5f and the height at the position of the black curve in fig. 5b are taken out and shown in fig. 5g respectively. It can be seen that although the morphology of the lesion fluctuates greatly, the temperature maintains high uniformity. Uniform and precise heating of specific locations on the skin is achieved.

Further by activating the patterned conformal LMPM, we developed conformal flexible devices on human skin and plant leaves (Fig. 5h – m). When the finger bends, straightens, etc., the wrinkle on the back of the finger changes, but the conformal flexible circuits always work stably, and the brightness of the yellow LED does not change (fig. 5h,i). Fig. 5j shows a conformal liquid metal-based strain sensor fabricated on the back of the thumb. There is no additional substrate between the sensor and the skin, so the interference from the mechanical properties of the substrate can be reduced. The resistive signal of the conformal sensor was continuously tested during continuous thumb movement (Fig. 5l). During the violent action of the thumb, the conformal film did not appear to fall off or break, which proves the adhesion ability of the liquid metal film to the human skin. LMPM is also transferred to the leaf and patterned to obtain strain sensors. Benefit from directly using the object under test as the substrate, reducing the constraints of the flexible substrate. During the entire experiment, the growth of the leaves is not disturbed by the sensor. Fig. 5m shows the data obtained by continuous monitoring for 20 days. In the first ten days, the leaf growth is slow due to cloudy weather, but in the next ten days under strong light conditions, the leaf grows rapidly, and the resistance of the strain sensor also increases quickly. In short, the conformal flexible electronic device based on LMPM can better reflect the real state of the measured object without the constraints of substrate.

## 3. Conclusion

In summary, Liquid metal particles are self-assembled into the large scale (>100 $cm^2$) monolayers at the air-water interface for realizing ultra-conformal E-tattoo. Benefiting from the shielding of interference from gravity in the MUS method and the assistance of Marangoni force, both nano to micro-sized (<500 μm) liquid metal particles can be self-assembled into monolayer in the air-water interface. The monolayer can be conformally transferred to silicon wafers, spherical surfaces, human skin, etc., and other complex surfaces. Based on the conformal liquid metal film, conformal flexible circuits and sensors are prepared on human skin, which shows good conformal ability and can perfectly adhere to complex structures such as wrinkles on the skin. Without the substrate limitation, the E-tattoo shows no interference with the measured object. Therefore, the user-friendly E-tattoo does not cause a foreign body sensation on the surface of the human body or plants, and will not hinder the movement of the human body and the growth of plants. The ultra-conformal liquid metal monolayer provides an essential tool for further research on conformal flexible circuits, wearable sensors, and other device applications.

**4. Experimental Section/Methods**

*Preparation of Liquid metal and liquid metal microparticle:* Gallium (99.99%; Beijing Founde Star Sci. & Technol. Co., Ltd), indium (99.995%; Beijing Founde Star Sci. & Technol. Co., Ltd), were mixed together in the ratio of 3:1 by mass. Then the mixture was heated and stirred for 30 min protected by nitrogen gas at 60 °C to obtain liquid-metal EGaIn (Ga75In25). The preparation of liquid metal particles utilizes a self-made high-pressure micro-pipe injection device. The liquid metal is injected into the dispensing syringe, and high-pressure nitrogen gas is introduced into the rear to force it through the micron glass tube. By controlling the diameter of the glass tube, monodisperse liquid metal particles of different sizes were obtained. The obtained liquid metal microparticles are cryopreserved to prevent interparticle agglomeration.

*Preparation of other micron materials:* Zinc sulfide microparticles were purchased from Shanghai keyan materials. The copper micro-particles and FeSiB micro-particles were purchased from Zhongnuo New Materials, and a metal screen was used to control their particle size uniformity.

*Preparation and conformal transferring of self-assembled particle films:* The following content takes liquid metal as an example, and the operations for other materials are the same. The liquid metal particles were stirred in alcohol for 10 minutes (ice bath, temperature below the melting point of the liquid metal). After standing for 5 minutes, the precipitate was taken and placed on a glass plate. When the liquid metal slurry on the glass plate contacts the air-water interface, the liquid metal particles automatically diffuse into the interface. After the film in the interface is saturated, the glass plate is removed to obtain a self-assembled microparticle film. Place the object below the water surface and slowly lift it up vertically. The self-assembled film at the interface will automatically attach to the surface of the object.

*Preparation of conformal devices on the human skin:* All wearable electronic experiments in this paper are carried out in the hands of the first author with his written consent. Firstly, paste a stickers-based mask on the surface of the finger. Then, insert the finger into the beaker that has been prepared with LMPM. The conformal liquid metal particle film attached to the surface of the finger can be obtained by slowly lifting the finger from water. After activating the film with a brush, the mask is removed, and finally a conformal liquid metal device on the skin surface is obtained. A multimeter is used for subsequent electrical tests.

**Supporting Information**

Supporting Information is available from the Wiley Online Library or from the author.

**Acknowledgements**

Special thanks to Professor Xuewen Wang from Northwestern Polytechnical University for his contribution in language polishing.

National Natural Foundation of China (552127803,51931011, 51971233, 62174165, M-0152, U20A6001, U1909215 and 52105286); External Cooperation Program of Chinese Academy of Sciences (174433KYSB20190038, 174433KYSB20200013); K.C. Wong Education Foundation (GJTD-2020-11); the Instrument Developing Project of the Chinese Academy of Sciences (YJKYYQ20200030); Chinese Academy of Sciences Youth Innovation Promotion Association (2018334); "Pioneer" and "Leading Goose" R&D Program of Zhejiang(2022C01032); Zhejiang Provincial Key R&D Program(2021C01183);Natural Science Foundation of Zhejiang Province (LD22E010002); Ningbo Scientific and Technological Innovation 2025 Major Project ( 2019B10127, 2020Z022); Zhejiang Provincial Basic Public Welfare Research Project (LGG20F010006); China Postdoctoral Science Foundation (2021M693249)

Received: ((will be filled in by the editorial staff))

Revised: ((will be filled in by the editorial staff))

Published online: ((will be filled in by the editorial staff))

**Competing Interests**

The Authors declare no Competing Financial or Non-Financial Interests

**Data Availability Statement**

The data that support the findings of this study are available from the corresponding author upon reasonable request.

**Author contributions:**

Conceptualization: F.L, RW.L, H.Y,

Methodology: F.L, W.L, H.Y,

Investigation: F.L, X.L, F.X, Z.H,

Visualization: F.L, S.L, J.L,

Funding acquisition: RW.L, Y.L, Y.W, J.S,

Project administration: Y.L, RW.L,

Supervision: RW.L, Y.L,

Writing – original draft: F.L,

Writing – review & editing: RW.L, Y.L, Y.W,

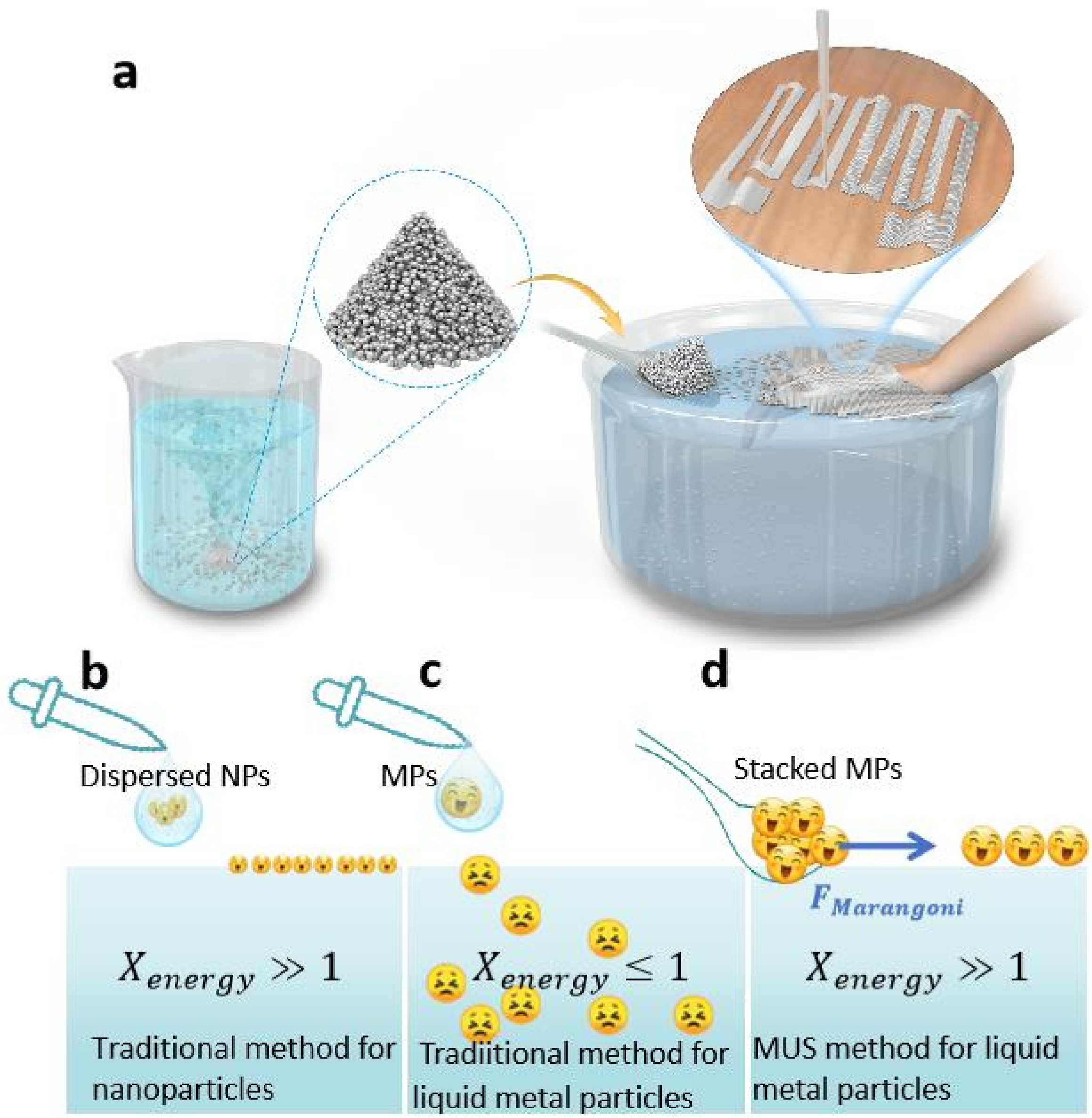

**Figure 1. Schematics of multi-size universal self-assembly (MUS) for liquid metal particle monolayer (LMPM)**. (a) The schematic diagram of fabricating liquid metal based conformal circuits. Mainly includes following steps: pre-dispersion of liquid metal particles, particle releasing at the interface, particles self-assembling into monolayer, and conformal transferring of LMPM. (b) The self-assembly of nanoparticles by traditional method. (c) Traditional method failed to assemble micro-scale liquid metal particles. (d) The self-assembly of liquid metal microparticles by MUS method.

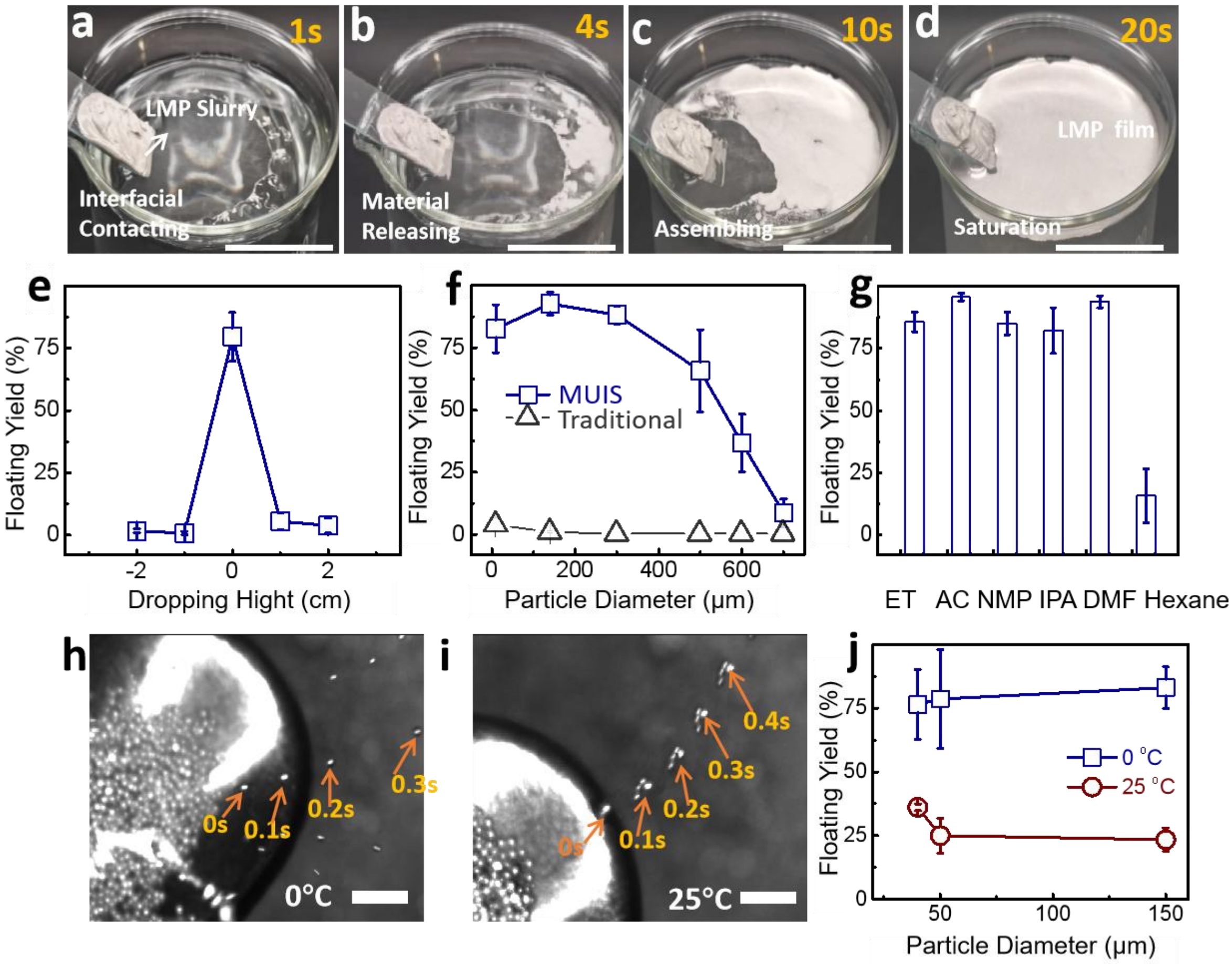

**Figure 2 Fabrication of liquid metal micoparticle monolayer.** (a-d) Photographs of different steps in MUS. Scale bars, 5 cm. (e) The floating yield according to dropping height of liquid metal slurry. (f) Floating yield of the traditional method (dropwize adding) and MUS method according to the diameter of liquid metal particles. (g) Floating yield of liquid metal particles based on different water soluble solvent (ethanol, acetone, n-methylpyrrolidone, n-dimethylformamide, isopropyl alcohol ) and water insoluble solvent (hexane). (h-i) Time-lapse microscopic photographs of the liquid metal particles in the "material releasing" process taken at 0°C and 25°C respectively. Scale bars, 500 μm.(j) Floating yield of liquid metal particles according to the temperature.

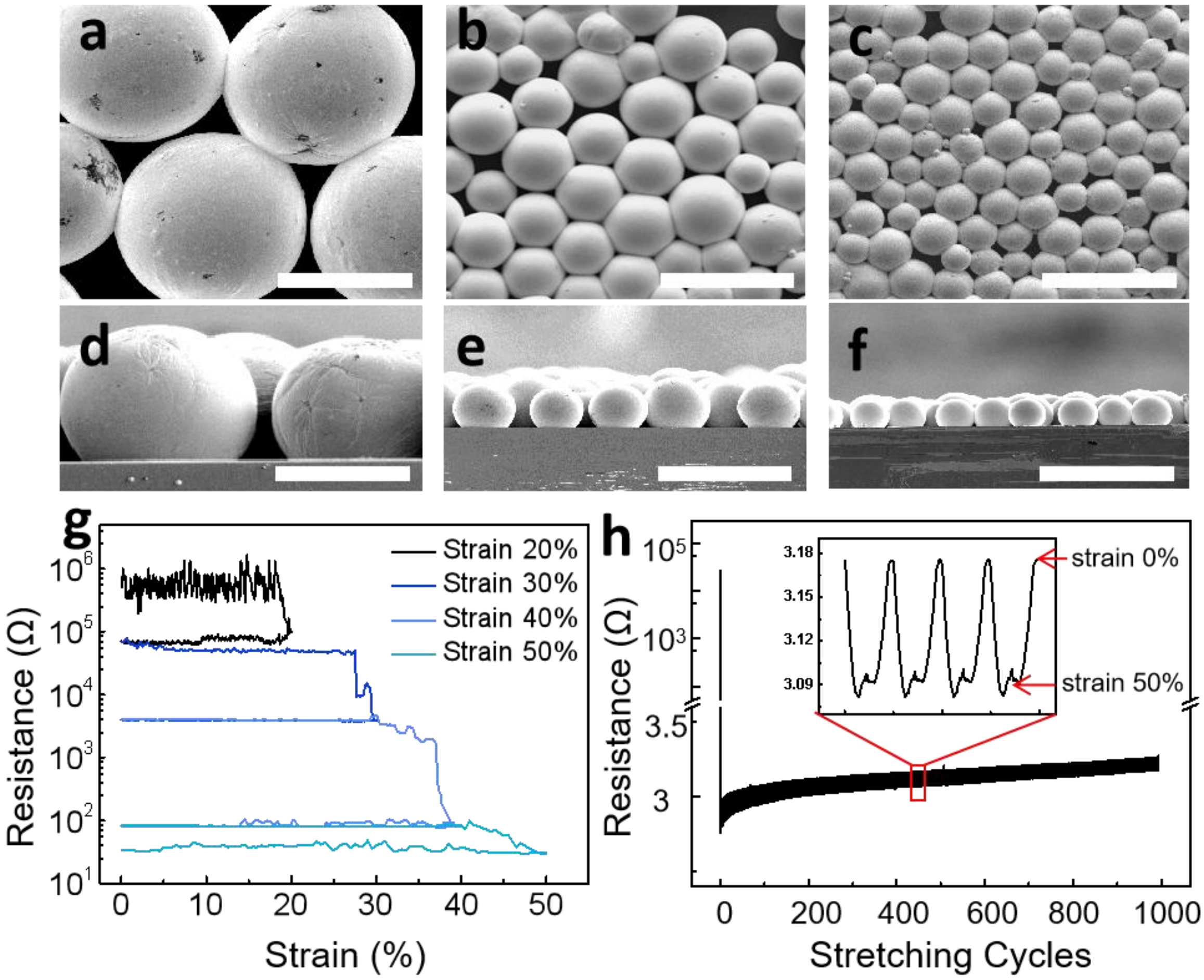

**Figure 3. The surface topography and electrical property of LMPM.** (a-c) SEM images of LMMM with different diameters (150 μm, 40 μm and 30 μm respectively). Scale bars, 100 μm. (d-f) Cross-sectional SEM images of the liquid metal monolayers in Fig. 3a-c. Scale bar: 100 μm. (g) Resistance of liquid metal monolyer (particles diameter: 30 μm) under different strain and the size of sample is 1cm×4cm. The maximum strain of the first cycle is 20%, then gradually increases by 10% until reaches 50%. (h) The resistance of LMPM as the function of stretching cycles.The size of sample is 1cm×4cm.

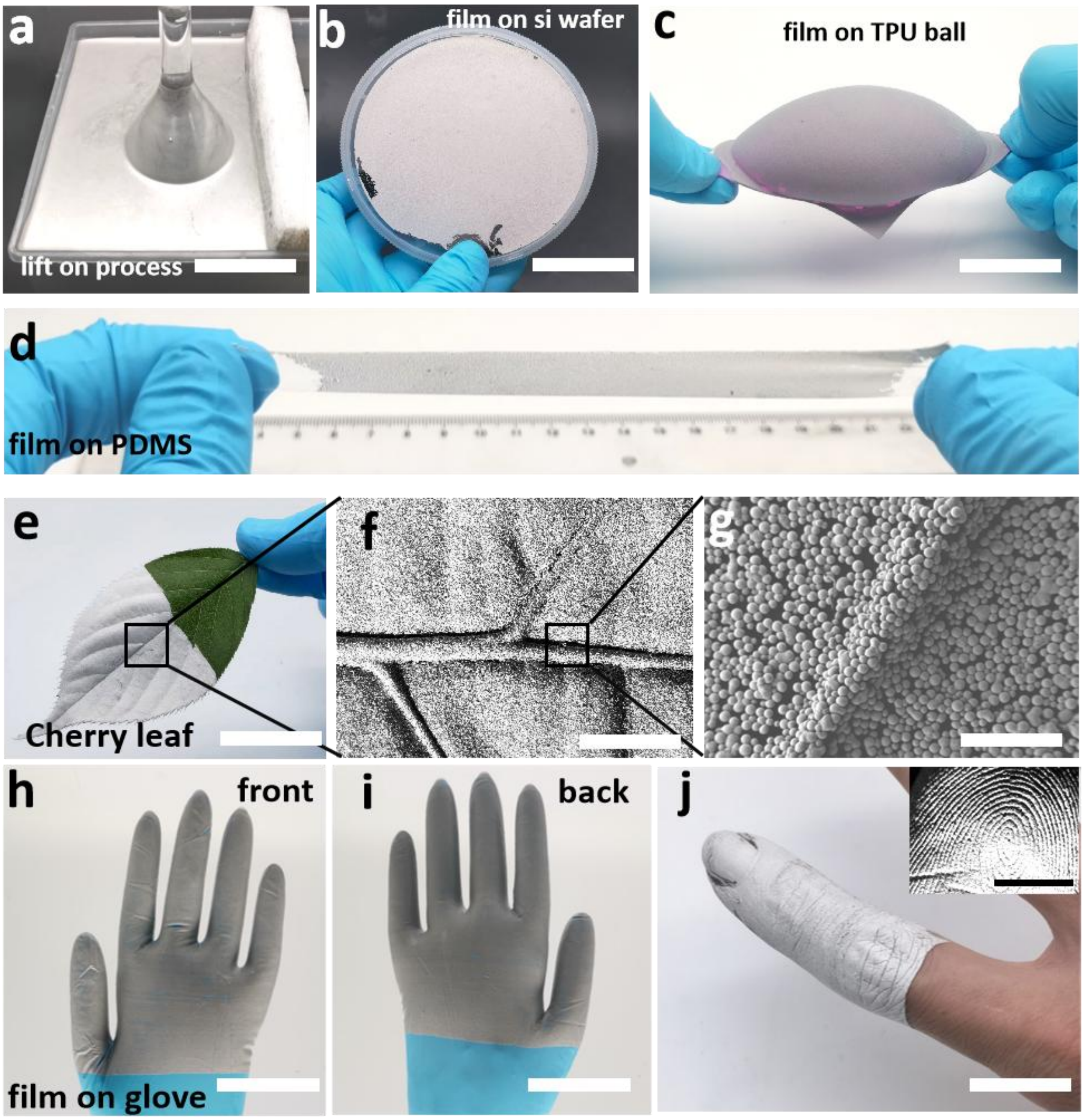

**Figure 4. The Conformal transfer of LMPM to various surfaces (flat, curved, complex).** (a) The Conformal transferring process of LMPM from the air-water interface to the surface of a glassware. Scale bars: 10 cm. (b-c) Photographs of LMPM transferred to silicon wafer and silicone hemisphere. Scale bars: 4 cm, 2 cm. (d) Photograph of the LMPM transferred to PDMS, which is being stretched with the strain of 100%. (e-g) Optical and SEM images of LMMM transferred to the surface of cherry leaf. Scale bars: 3 cm, 2 mm, 500 µm. (h-i) Optical photographs of self-assembled liquid metal films on the surface of gloves, front and back, respectively. Scale bars: 5 cm. (j) Optical photograph of LMPM on the index finger. Scale bar: 2 cm. Inset: Microscopic image of the film on the fingerprint. Scale bar: 500 µm.

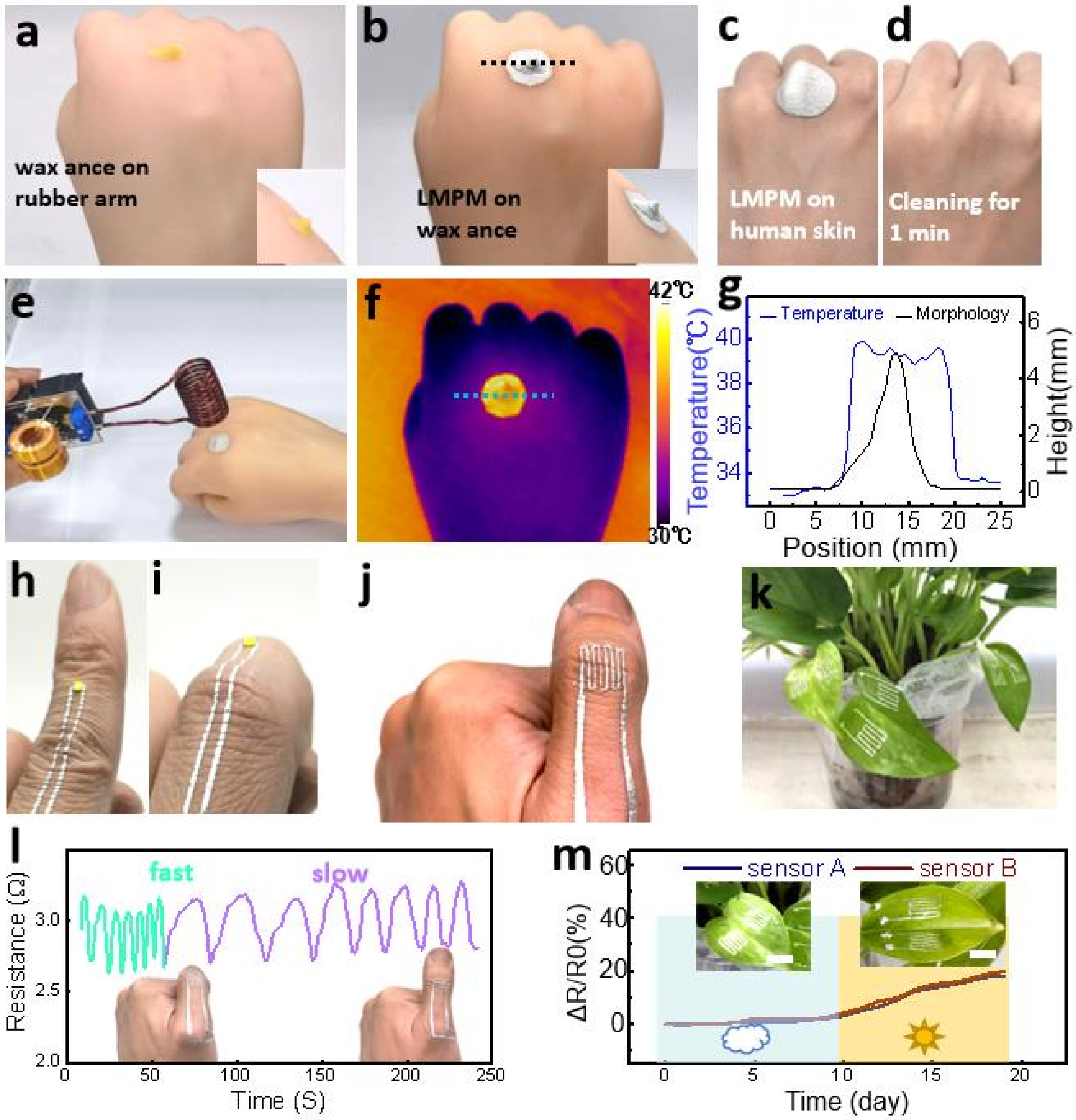

**Figure 5** liquid metal E-tattoo on human skin as wireless conformal heater and conformal flexible circuits and sensors. (a) Wax is used to construct a raised structure on the surface of the silicone prosthesis, which simulates the structure of human skin lesions. (b) The LMPM is transferred to the "lesion site" to achieve conformal coverage of this structure. (c-d) LMPM transferred to the joints of human skin, which can be completely removed after a one-minute soap washing step, leaving the skin clean without residue. (e) Non-contact heating of conformal LMPM is achieved using a commercial high-frequency low voltage induction heater. The temperature is controlled by controlling the time and distance between the induction coil and the film. (f) Infrared image of the prosthetic skin after ten seconds of heating. (g)The contrast of height and temperature where the lesion is located. Corresponding to the position of the black dotted line in fig. 5b and the position of the blue dotted line in fig. 5f, respectively. (h,i) The conformal liquid metal circuit located on the back of the finger.(j) The conformal strain sensor on the back of the thumb. (k) The conformal strain sensor on the leaf of epipremnum aureum. (l)The curve of strain sensor′s resistance versus time. The strain sensor is located at the back of thumb which makes the flexing-stretching motion. (l) The resistance change of strain sensor during leaf growth. The plants were in the shade for the first ten days, and were exposed to sunlight or artificial light for the next ten days.

Conformal liquid metal particles monolayer is proposed by the multi-size universal interfacial self-assembly method. By optimizing the traditional interfacial self-assembly through a z-axis undisturbed particle releasing strategy, both nano and micro-sized liquid metal particles can be fabricated into monolayer. Furthermore, the self-assembled film can be conformal transferred to complex surfaces for conformal electronics.

F. Li, W. Lei, Y. Wang, X. Lu, S. Li, F. Xu, Z. He, J. Liu, H. Yang, Y. Wu, J. Shang, Y. Liu*, R-W. Li*

**Ultra-conformable Liquid Metal Particle Monolayer on Air/water Interface for Substrate-free E-tattoo**

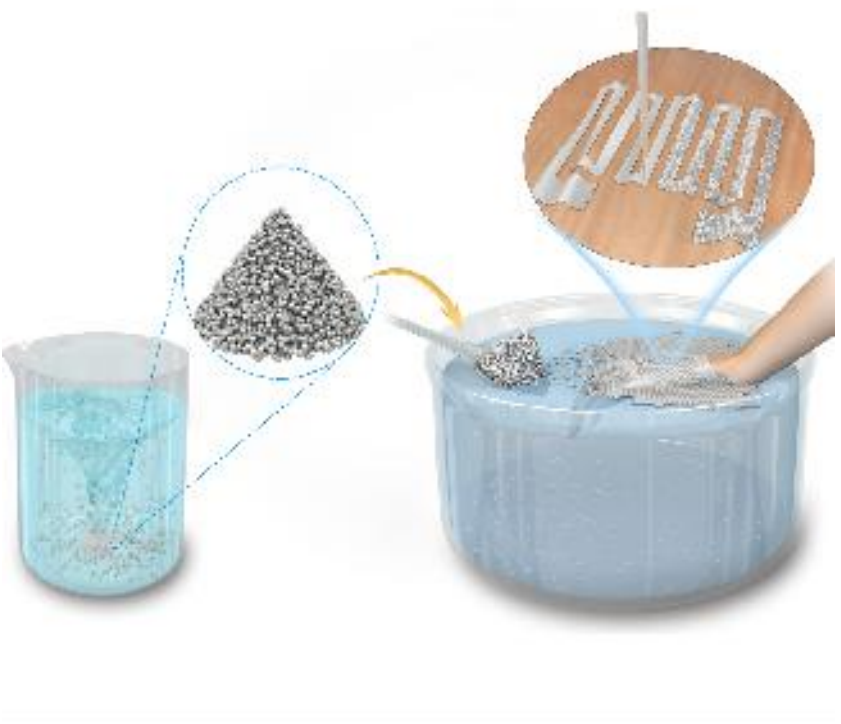

Supporting Information

## Ultra-conformable Liquid Metal Particle Monolayer on Air/water Interface for Substrate-free E-tattoo

*Fali Li, Wenjuan Lei, Yuwei Wang, Xingjian Lu, Shengbin Li, Feng Xu, Zidong He, Jinyun Liu, Huali Yang, Yuanzhao Wu, Jie Shang, Yiwei Liu*, Run-Wei Li**

**This PDF file includes:**

Figs. S1 to S6

Captions for Movies S1 to S3

**Other Supplementary Materials for this manuscript include the following:**

Movies S1 to S3

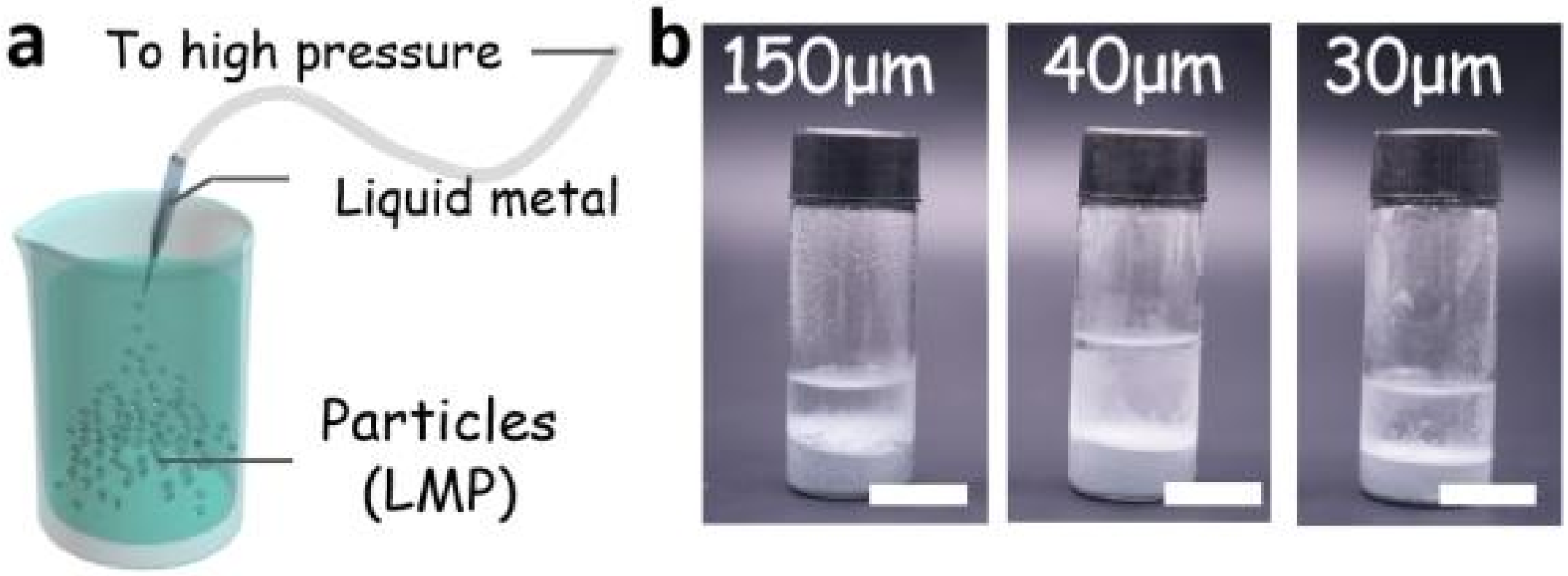

**Figure S1. Preparation of mono-dispersed liquid metal particles.** (a) The liquid metal becomes particles when passes through glass micro-tubes at high pressure, and mono-disperse liquid metal particles of different sizes are obtained by By controlling the diameter of the micro-needles. (b)The optical images of liquid metal particles with different diameters in glass bottle.

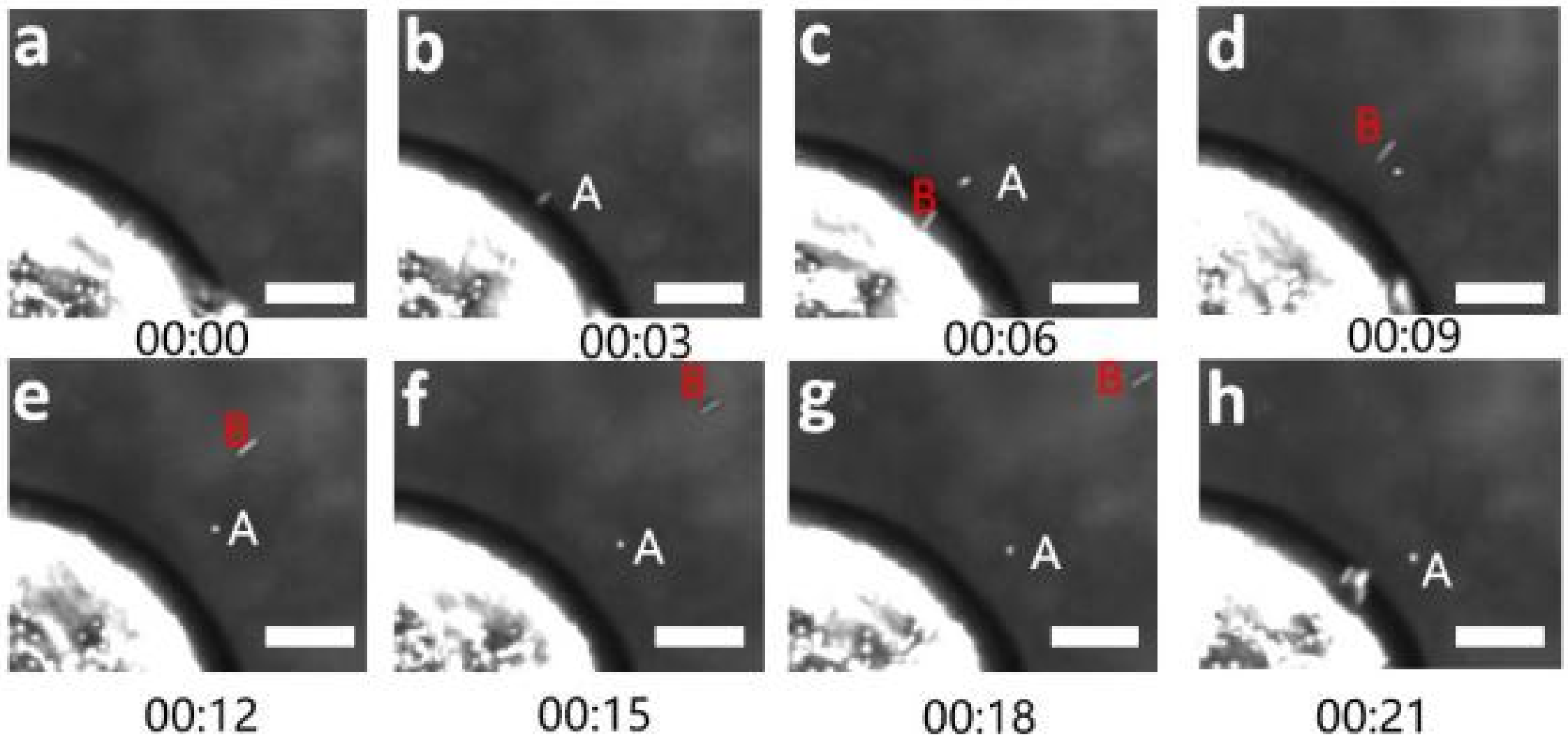

**Figure S2. Photograph of the material releasing process which is mediated by the Marangoni force.** Photos are taken by a high-speed microscopic camera. After particle A is launched from the spoon, it decelerates rapidly and sinks into the water. In contrast, particle B always keeps moving forward at a relative constant speed. Scale bars, 500 μm.

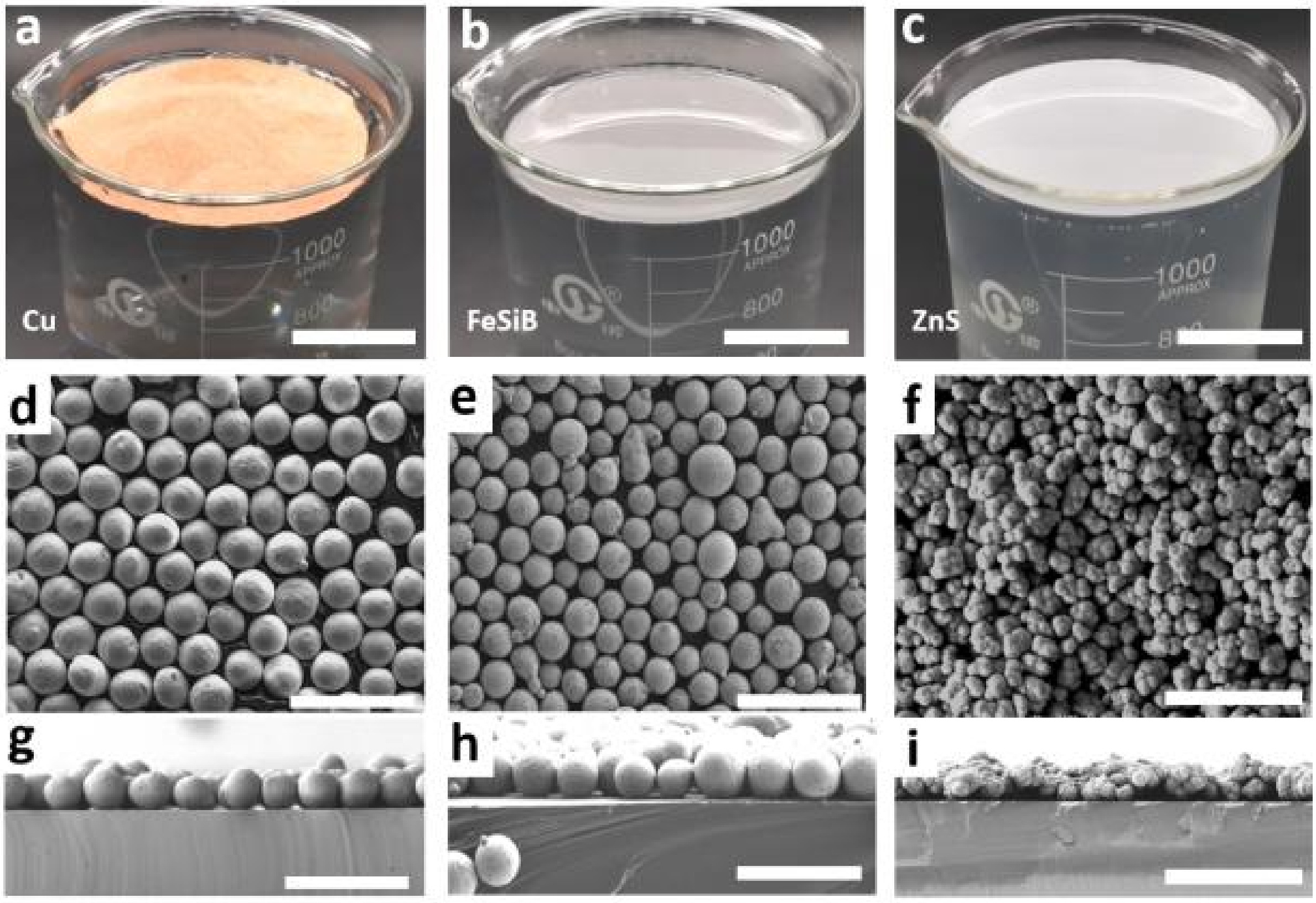

**Figure S3. The universality of size-universal self-assembly**. (a-c) Photographs of different materials assembled at air-water interface. Left to right: Cu MPs, FeSiB MPs, ZnS MPs, respectively. Scale bars, 5 cm (d-i) Scanning electron microscope images and cross-sections of different microparticle monolayers. Left to right: Cu MPs, FeSiB MPs, ZnS MPs, respectively. Scale bars: 500 µm, 100 µm, 20 µm

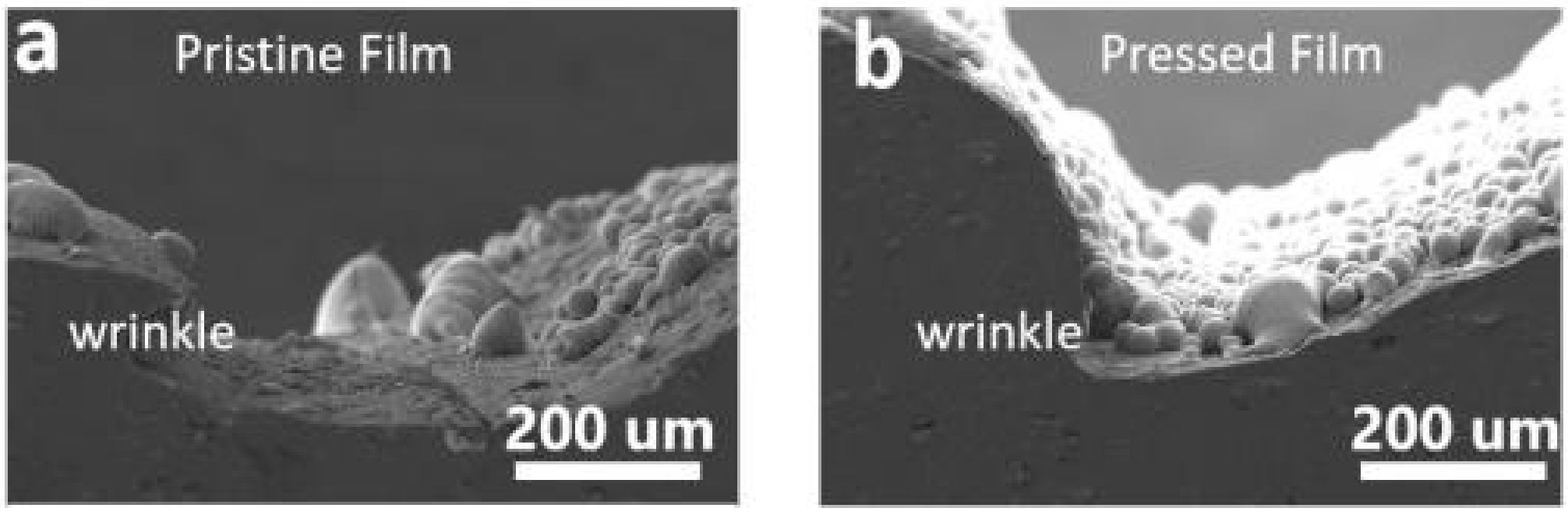

**Figure S4. SEM images of pristine and compressed films transferred to wrinkled surfaces**. (a) The LMPM is directly transferred to the uneven surface. SEM images show that it is difficult to achieve fully coverage in areas with excessive surface undulation. (b) After the liquid metal self-assembled film is pre-compressed, the complex wrinkle surface can be effectively covered.

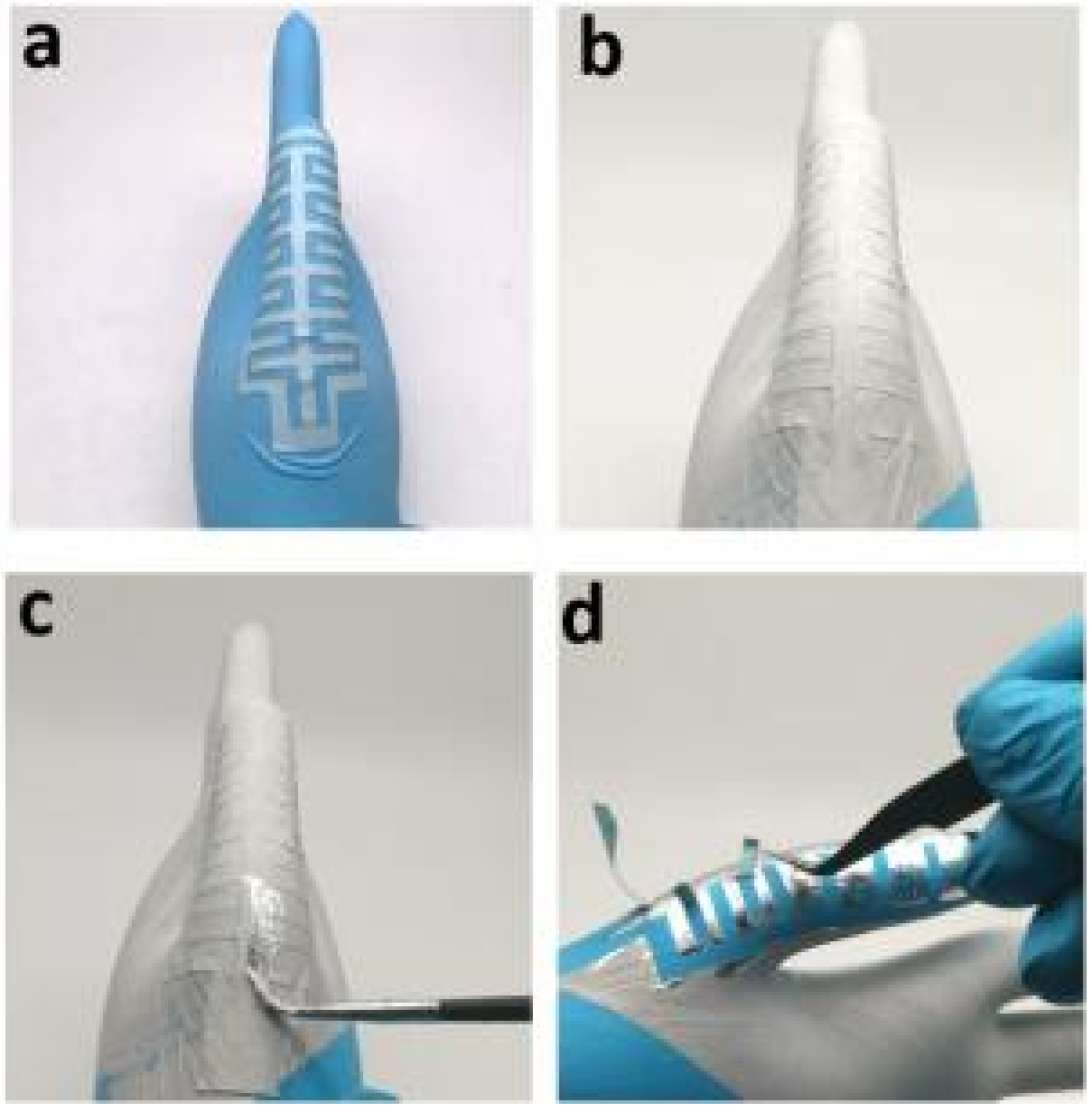

**Figure S5. The process of fabricating conformal liquid metal circuits on the surface of gloves.**(a) Attach the paper reticle to the surface of the glove. (b) Conformal transferring of LMPM to the glove surface. (c) The particle film is processed with a brush to realize a continuous conductive path. (d) Use tweezers to remove the reticle.

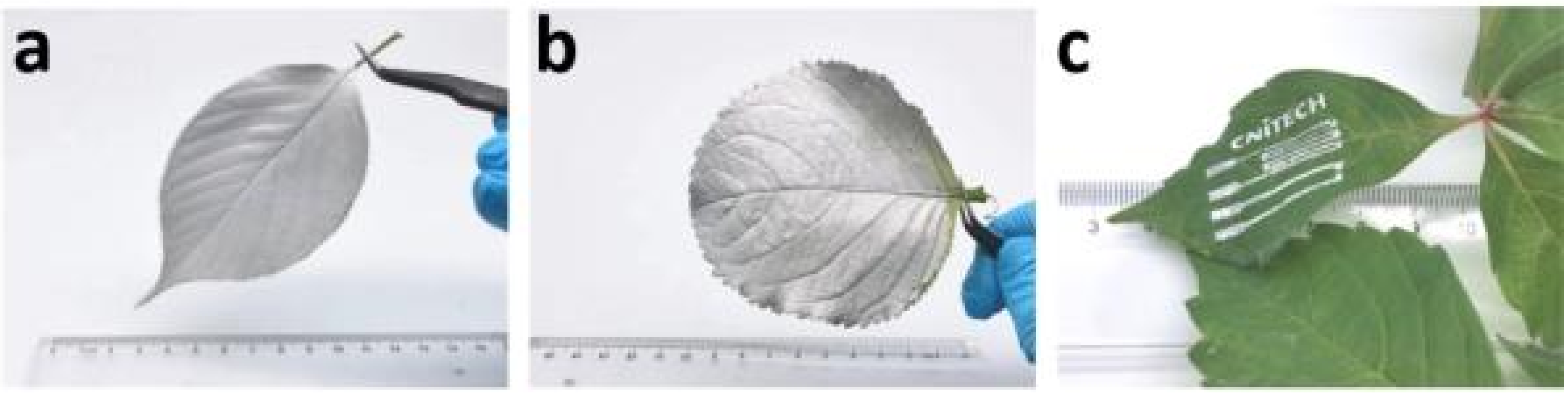

**Figure S6. The LMPM transferred to the surface of different plant leaves.** (a) Cherry leaf whose entire surface is covered with LMPM. (b) hydrangea leaf whose entire surface is covered with LMPM. (c) The pattern of LMPM on the leaf of creeper.

**Movie S1.**

The process of preparing liquid metal particle film by MUS method. It only takes 20S to fabricate a film with diameter of 10 cm. Then, the beaker was tapped to form ripples on the water surface, and the film remained stable.

**Movie S2.**

The process of conformal transfer of the film to the skin of the finger. Insert a finger into the obtained film, and as the finger is lifted up and away from the water surface, the liquid metal particle film is automatically and conformally attached to the surface of the finger, and the details such as the fingerprint folds of the skin are kept.

**Movie S3.**

Liquid metal conformal circuit located on the skin. The circuit keeps working stably all the time during the movement of human fingers, and keeps supplying power to an LED.